\documentclass[aps,pra,twocolumn,showpacs,superscriptaddress,groupedaddress]{revtex4} 

\usepackage[T1]{fontenc}
\usepackage{amsmath,amsfonts,amssymb,amsthm,bbm}
\usepackage{graphicx}
\usepackage{color}

\allowdisplaybreaks

\input{Qcircuit}

\begin{document}

\title{Computational complexity of non-equilibrium steady states of quantum spin chains}
\author{Ugo Marzolino}
\author{Tomaz Prosen}
\affiliation{Univerza v Ljubljani, SI-1000 Ljubljana, Slovenija}

\date{\today}

\begin{abstract}
We study non-equilibrium steady states (NESS) of spin chains with boundary Markovian dissipation from the computational complexity point of view. We focus on XX chains whose NESS are matrix product operators (MPO), i.e. with coefficients of a tensor operator basis described by transition amplitudes in an auxiliary space. Encoding quantum algorithms in the auxiliary space, we show that estimating expectations of operators, being local in the sense that each acts on disjoint sets of few spins covering all the system, provides the answers of problems at least as hard as, and believed by many computer scientists to be much harder than, those solved by quantum computers. We draw conclusions on the hardness of the above estimations.
\end{abstract}

\maketitle

\section{Introduction}

Integrable systems offer powerful tools to grasp important physical properties of low dimensional strongly correlated systems \cite{Korepin1993,Giamarchi2003}. Examples are the Heisenberg XXZ model or the XY Hamiltonian, solved by Bethe Ansatz and fermionization respectively, that describe magnetic phenomena \cite{Schollwock2004} and spin chain materials \cite{Sologubenko2007}. Moreover, quantum information and complexity offer elaborated characterisations of properties of quantum correlated systems \cite{Kempe2006,Oliveira2008,Schuch2009,Whitfield2013,Childs2016,Mehraban2015}. In particular, computational complexity theory \cite{AroraBarak,DuKo,Watrous2009} studies the complexity of certain mathematical problems as compared to known hard ones.

Recently, Bethe Ansatz techniques were extended to find the non-equilibrium steady state (NESS) of a XXZ spin chain with boundary dissipation (see \cite{Prosen2014,Prosen2015} for a review). This NESS is a matrix product operator (MPO), namely a density matrix with coefficients of a given operator tensor basis being transition amplitudes in an auxiliary Hilbert space. This result was used to lower bound the Drude weight, proving ballistic spin transport \cite{Prosen2011a,Ilievski2013}. Moreover, NESSs were investigated concerning intertwined topics, such as geometry of states, linear responce theory, non-equilibrium phase transitions, and quantum metrology \cite{Prosen2008,Banchi2014,Marzolino2014,CamposVenuti2015,Albert2015}. MPO were also used to numerically simulate the time-evolution of states under dissipation \cite{Verstraete2004,Zwolak2004}, and for a numerical variational approach to NESS \cite{Cui2015}.

Pure state versions of MPO, i.e. matrix product states (MPS) and projected entangled pair states (PEPS), have many applications, such as fixed points of the density matrix renormalization group \cite{Ostlund1995,Schollwock2005}, the efficient simulation of one-dimensional many-body systems \cite{Vidal2004} and quantum computations with slight entanglement \cite{Vidal2003}, even though the latter ambiguously depends on the specific entanglement measure \cite{VandenNest2013}, and measurement based quantum computation \cite{Raussendorf2003,Childs2005,Gross2007-2}. PEPS were also studied from the perspective of computational complexity, proving that both creating them and computing their local expectation values are very hard problems (\textsf{PP} and \textsf{\#{}P} respectively) \cite{Schuch2007}, at least as hard as more famous \textsf{NP-complete} problems \cite{Verstraete2006}. Furthermore, decision problems for ground states of many Hamiltonians are \textsf{QMA-complete} \cite{Kempe2006,Oliveira2008,Schuch2009,Whitfield2013,Childs2016}, the quantum analogue of \textsf{NP-complete}.

In this paper, we characterise the computational complexity of estimating expectation values of local operators in NESS described by MPO, by mapping quantum circuits, given by sequences of elementary operations, called gates, to its auxiliary space. Deciding whether a MPO is positive, thus a density operator, is a computationally hard problem (\textsf{NP-hard}) for finite size and even undecidable for infinitely large systems \cite{Kliesch2014}. Thus, applications of MPO should be worked out with concrete MPO known to be density operators, like NESS. We find that the estimation of certain local operators of the NESS of spin chains enables us to solve extremely hard problems.

Our result suggests that the number of required measurements is exponentially large. If not, \textsf{PP} problems would be efficiently solved by quantum computers, using properties of the latter, with tremendous consequences in computational complexity: \textsf{PP} problems would not increase allowing magical subroutines that efficiently and deterministically solve \textsf{PP} problems; \textsf{PP} problems, and so those efficiently solved by quantum computers, would contain the polynomial hierarchy \textsf{PH}, i.e. the union of a, conjectured infinite, hierarchy of problems that generalise \textsf{P} and \textsf{NP} \cite{AroraBarak,DuKo}. These conclusions suggest the enormous difficulty of the above estimations for both experiments and computations, and extend applications of computational complexity to non-equilibrium physics. Our result also provides new properties of \textsf{PP} problems which are useful tools for further proofs of classical and quantum complexity. Finally, the present approach strengthens the connection between NESS and quantum information, already established through metrological applications \cite{Marzolino2014}.

\begin{figure}[htbp]
\includegraphics[width=\columnwidth]{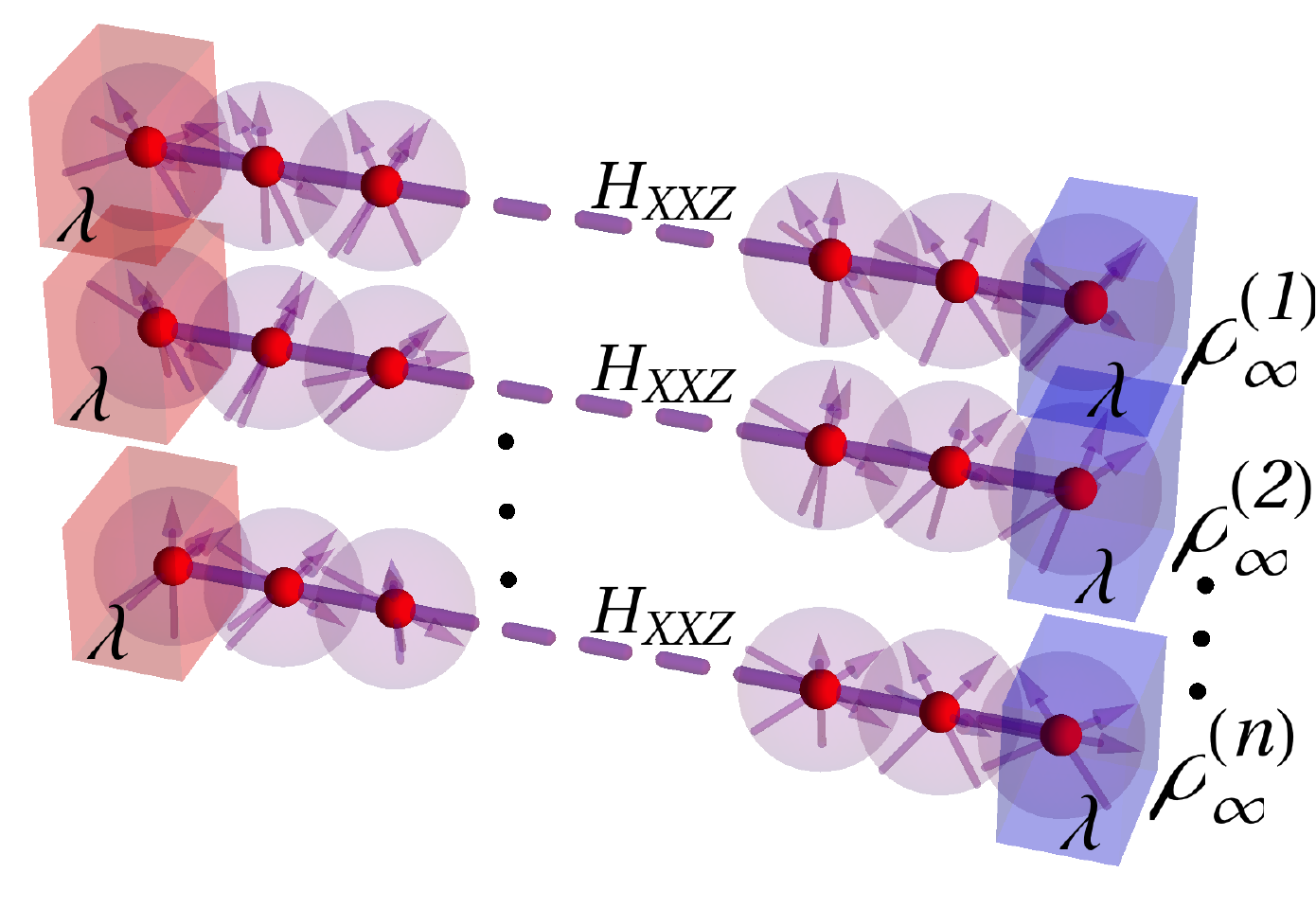}
\caption{(Color online) NESS of XXZ spin chains with boundary noise.}
\label{chains}
\end{figure}

\section{The systems}

We study the computational complexity of $n$ chains with $L$ spins $1/2$, as in Fig. \ref{chains}, in the state

\begin{equation} \label{NESS}
\rho_\infty=\prod_{k=1}^n\rho_\infty^{(k)}, \qquad \textnormal{where} \qquad \rho_\infty^{(k)}=\frac{S_k S_k^\dag}{\textnormal{Tr}\left(S_k S_k^\dag\right)}
\end{equation}
is NESS of the $k$-th chain under the following dynamics

\begin{eqnarray}
\frac{d}{dt}\rho_t^{(k)} & = & -i\left[h\sum_{j=1}^L\sigma_{k,j}^z+H_\textnormal{XXZ}^{(k)},\rho_t^{(k)}\right] \nonumber \\
& & +\lambda\sum_{l=1}^2\left(\mathcal{L}_l\rho_t^{(k)}\mathcal{L}_l^\dag-\frac{1}{2}\left\{\mathcal{L}_l^\dag\mathcal{L}_l,\rho_t^{(k)}\right\}\right).
\label{master.eq}
\end{eqnarray}
The XXZ Hamiltonian is

\begin{equation}
H_\textnormal{XXZ}^{(k)}=\sum_{j=1}^{L-1}(\sigma_{k,j}^x\sigma_{k,j+1}^x+\sigma_{k,j}^y\sigma_{k,j+1}^y+
\Delta\sigma_{k,j}^z\sigma_{k,j+1}^z),
\end{equation}
Lindblad noise driving channels are $\mathcal{L}_1=\sigma_{k,L}^+$ and $\mathcal{L}_2=\sigma_{k,1}^-$, and $\sigma_{k,j}^{x,y,z}$ ($\sigma_{k,j}^\pm=(\sigma_{k,j}^x\pm i\sigma_{k,j}^y)/2$) are the Pauli matrices of the $j$-th spin in the $k$-th chain \footnote{$\sigma_{k,j}^\alpha$ is the tensor product of the Pauli matrix $\sigma^\alpha$ on the $j$-th spin in the $k$-th chain and the identity on the other spins. We reserve the tensor product notation only for the product of matrices acting on auxiliary spaces. We keep the chain label on operators acting on the auxiliary space of the corresponding chain even though unnecessary in the tensor product notation, in order to stress the correspondence between the auxiliary and the physical spaces.}. The exact NESS of \eqref{master.eq} was computed in \cite{Prosen2011b}, and enjoys a MPO structure using an auxiliary Hilbert space spanned by the orthonormal bases $\{|0_k\rangle,|1_k\rangle,|2_k\rangle,\dots,|\lfloor\frac{L}{2}\rfloor_k\rangle\}$:

\begin{equation} \label{MPO}
S_k=\sum_{\substack{\{s_{k,1},\dots,s_{k,L}\}\\\in\{0,+,-\}^L}}\langle 0_k|\prod_{j=L}^1 A_{s_{k,j}}^{(k)}|0_k\rangle\prod_{j=L}^1\sigma_{j,k}^{s_{k,j}},
\end{equation}
with tridiagonal matrices $A_{s}^{(k)}$ on the auxiliary Hilbert spaces. The dimension of the auxiliary space is at most $1+\lfloor\frac{L}{2}\rfloor$, but can be smaller and independent of $L$ for certain values of $\Delta$ \cite{Prosen2011b}. We now consider the XX model, i.e. $\Delta=0$, which corresponds to the minimal (two) dimension of the auxiliary space for any number $L$ of spins in a single chain:

\begin{eqnarray}
A_0^{(k)} & = & |0_k\rangle\langle 0_k|+\frac{i\lambda}{4} \, |1_k\rangle\langle 1_k|, \nonumber \\
A_+^{(k)} & = & \frac{i\lambda}{2} \, |0_k\rangle\langle1_k|, \nonumber \\
A_-^{(k)} & = & |1_k\rangle\langle0_k|. \label{A}
\end{eqnarray}
Using equation \eqref{MPO} into \eqref{NESS}, the NESS is explicitly written as another MPO:

\begin{equation} \label{NESS1}
\rho_\infty^{(k)}= \!\!\!\!\!\!\! \sum_{\substack{\{s_{k,1},\dots,s_{k,L}\}\\\in\{0,z,+,-\}^L}} \!\!\!\!\!\!\! \frac{\langle0_{+k}|\langle0_{-k}|\prod_{j=L}^1\mathbb{A}_{s_{k,j}}^{(k)}|0_{+k}\rangle|0_{-k}\rangle}{\prod_{k=1}^n\textnormal{Tr}\left(S_k S_k^\dag\right)}\bigotimes_{j=L}^1\sigma_{j,k}^{s_{k,j}},
\end{equation}
where we have defined the matrices

\begin{eqnarray}
\mathbb{A}_0^{(k)} & = & A_0^{(k)}\otimes\bar A_0^{(k)}+\frac{1}{2}\sum_{\epsilon=\pm}A_\epsilon^{(k)}\otimes\bar A_\epsilon^{(k)}, \nonumber \\
\mathbb{A}_z^{(k)} & = & \frac{1}{2}\Big(A_+^{(k)}\otimes\bar A_+^{(k)}-A_-^{(k)}\otimes\bar A_-^{(k)}\Big), \nonumber \\
\mathbb{A}_\pm^{(k)} & = & A_\pm^{(k)}\otimes\bar A_0^{(k)}+A_0^{(k)}\otimes\bar A_\mp^{(k)}. \label{AA}
\end{eqnarray}
Note that the auxiliary Hilbert space in \eqref{NESS1} is doubled with respect to the auxiliary space in \eqref{MPO}, due to the product $S_k S_k^\dag$ in \eqref{NESS}, see also (\ref{AA},\ref{prod.seq},\ref{loc.seq}).

Markovian master equations with local Lindblad operators, i.e. each environment interacting with a single particle as in \eqref{master.eq}, can be derived from microscopic models of system-environment interaction using the singular coupling limit \cite{BreuerPetruccione,Benatti2005}, or with a Davis's weak coupling approach when the magnetic field $h$ dominates the $XXZ$ interaction \cite{Wichterich2007}. The strength of the magnetic field $h$ does not alter the NESS, because the local Hamiltonian generator $-i\big[h\sum_{j=1}^L\sigma_{k,j}^z,\cdot\big]$ commutes with the other terms in \eqref{master.eq}, namely $-i\big[H_\textnormal{XXZ}^{(k)},\cdot\big]$ and $\lambda\big(\mathcal{L}_l\cdot\mathcal{L}_l^\dag-\frac{1}{2}\big\{\mathcal{L}_l^\dag\mathcal{L}_l,\cdot\big\}$. Markovian approximations hold for large time \cite{BreuerPetruccione,Benatti2005}. Moreover, the spectral gap of the Liouvillian \eqref{master.eq}, i.e. the smallest non-vanishing real part of its eigenvalues, goes to zero as one divided by a polynomial of the chain length \cite{Prosen2008}. Thus, only polynomial time complexity overhead is required to prepare the NESS.

Thus, a time that scales polynomially in the system size is enough to prepare the NESS. This means that the NESS preparation has polynomial time complexity, and thus is efficient in computational complexity terms.

\section{Quantum computation in the auxiliary space}

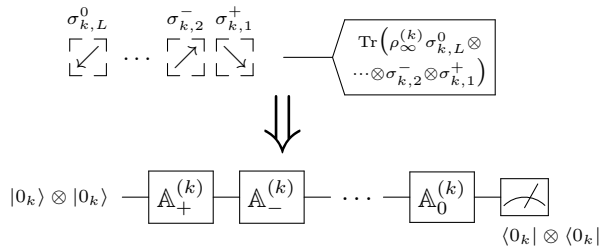
\begin{figure}[htbp]
\[
\Qcircuit @C=2em @R=0em {
\scriptsize{\textnormal{$\sigma_{k,L}^0$}} && \scriptsize{\textnormal{$\sigma_{k,2}^-$}} & \scriptsize{\textnormal{$\sigma_{k,1}^+$}} \\
\swarrow & \dots & \nearrow & \searrow && \measuretab{\substack{\textnormal{Tr}\big(\rho_\infty^{(k)}
\sigma_{k,L}^0\otimes \\ \dots\otimes\sigma_{k,2}^-\otimes\sigma_{k,1}^+\big)}}
\gategroup{2}{1}{2}{1}{1.5em}{--}
\gategroup{2}{3}{2}{3}{1.5em}{--}
\gategroup{2}{4}{2}{4}{1.5em}{--}
}
\]
\vspace{-25pt}
\Huge
\[
\Downarrow
\]
\normalsize
\vspace{-15pt}
\[
\qquad\qquad
\Qcircuit @C=1em @R=0.5em {
\lstick{\scriptsize{\textnormal{$\ket{0_k}\otimes\ket{0_k}$}}} & \gate{\mathbb{A}_+^{(k)}} & \gate{\mathbb{A}_-^{(k)}} & \qw & \dots && \gate{\mathbb{A}_0^{(k)}} & \meter \\
&&&&&&& \qquad \scriptsize{\textnormal{$\bra{0_k}\otimes\bra{0_k}$}}
}
\]
\caption{Picture of the encoder and the encoding \eqref{prod.seq} of single qubit operations in the virtual space of a spin chain in NESS.}
\label{1qubit.circ}
\end{figure}

We map sequences of quantum circuit gates into auxiliary spaces of spin chains through sequences of matrices $\mathbb{A}_{s}^{(k)}$. As a pedagogical example, consider the following expectation

\begin{equation} \label{prod.seq}
\frac{\textnormal{Tr}\left(\rho_\infty^{(k)}\prod_{j=L}^1(\sigma_{k,j}^{s_{k,j}})^\dag\right)}{\prod_{j=L}^1 f_{s_{k,j}}}=\frac{\langle 0|\langle 0|\prod_{j=L}^1\mathbb{A}_{s_{k,j}}^{(k)}|0\rangle|0\rangle}{\textnormal{Tr}\left(S_k S_k^\dag\right)},
\end{equation}
where $f_{s_{k,j}}=\textnormal{Tr}\big(\sigma_{k,j}^{s_{k,j}}(\sigma_{k,j}^{s_{k,j}})^\dag\big)$ is the Hilbert-Schmidt norm of $\sigma_{k,j}^{s_{k,j}}$ ($f_{0,z}=2$, $f_{\pm}=1$). Eq. \eqref{prod.seq} encodes a sequence of operations in the auxiliary space of a single spin chain, depicted in Fig. \ref{1qubit.circ}.

\begin{figure}[htbp]
\[
\hspace{30pt}
\Qcircuit @C=2em @R=-2.2em @!R {\scriptsize{\textnormal{$\mathcal{E}(\texttt{Fin}_\texttt{1})$}} & \quad\quad \scriptsize{\textnormal{$\mathcal{E}(\texttt{X}_\texttt{1})$}} & & \quad\quad \scriptsize{\textnormal{$\mathcal{E}(\texttt{C-Z}_{\texttt{1,2}})$}} & & \,\,\, \scriptsize{\textnormal{$\mathcal{E}(\texttt{In}_\texttt{1})$}} \\
& & & & & & & & & && \\
\swarrow & \nearrow & \nwarrow & \nearrow & \swarrow & \nearrow \\
& & & & & && \measuretab{\substack{\textnormal{Tr}\big(\rho_\infty\mathcal{E}(\texttt{Fin}_\texttt{1}) \\ \otimes\dots\otimes\mathcal{E}(\texttt{In}_\texttt{2})\big)}} \\
\searrow & \nwarrow & \searrow & \nwarrow & \nearrow & \searrow \\
& & & & & & & && \\
\scriptsize{\textnormal{$\mathcal{E}(\texttt{Fin}_\texttt{2})$}} & \quad\quad \scriptsize{\textnormal{$\mathcal{E}(\texttt{H}_\texttt{2})$}} & & & & \,\,\, \scriptsize{\textnormal{$\mathcal{E}(\texttt{In}_\texttt{2})$}}
\gategroup{3}{6}{3}{6}{1.5em}{--}
\gategroup{5}{6}{5}{6}{1.5em}{--}
\gategroup{3}{4}{5}{5}{1.5em}{--}
\gategroup{3}{2}{3}{3}{1.5em}{--}
\gategroup{5}{2}{5}{3}{1.5em}{--}
\gategroup{3}{1}{3}{1}{1.5em}{--}
\gategroup{5}{1}{5}{1}{1.5em}{--}
}
\]
\vspace{-10pt}
\Huge
\[
\Downarrow
\]
\normalsize
\vspace{-20pt}
\[
\Qcircuit @C=0.5em @R=1em {
\lstick{\scriptsize{\textnormal{$\ket{0_1}\otimes\ket{0_1}$}}} & \measure{\texttt{In}} & \qw & \ustick{\scriptsize{\textnormal{\!\!\! $\frac{|\texttt{0}\rangle+|\texttt{1}\rangle}{\sqrt{2}}$}}} \qw & \qw & \ctrl{1} & \gate{\texttt{X}} & \qw & \qw & \qw & \qw & \qw & \qw & \qw & \qw & \measure{\texttt{Fin}} & \meter & \rstick{\!\!\!\! \scriptsize{\textnormal{$\bra{0_1}\otimes\bra{0_1}$}}} \\
\lstick{\scriptsize{\textnormal{$\ket{0_2}\otimes\ket{0_2}$}}} & \measure{\texttt{In}} & \qw & \dstick{\scriptsize{\textnormal{\!\!\! $\frac{|\texttt{0}\rangle+|\texttt{1}\rangle}{\sqrt{2}}$}}} \qw & \qw & \gate{Z} & \gate{\texttt{H}} & \qw & \qw & \qw & \qw & \ustick{\scriptsize{\textnormal{$\frac{|\texttt{0}\rangle\otimes|\texttt{1}\rangle+|\texttt{1}\rangle\otimes|\texttt{0}\rangle}{\sqrt{2}}$}}} \qw & \qw & \qw & \qw & \measure{\texttt{Fin}} & \meter & \rstick{\!\!\!\! \scriptsize{\textnormal{$\bra{0_2}\otimes\bra{0_2}$}}} \\
}
\]
\caption{Picture of the encoder and the encoding of a simple circuit in the virtual spaces of spin chains in NESS.}
\label{ex.circ}
\end{figure}
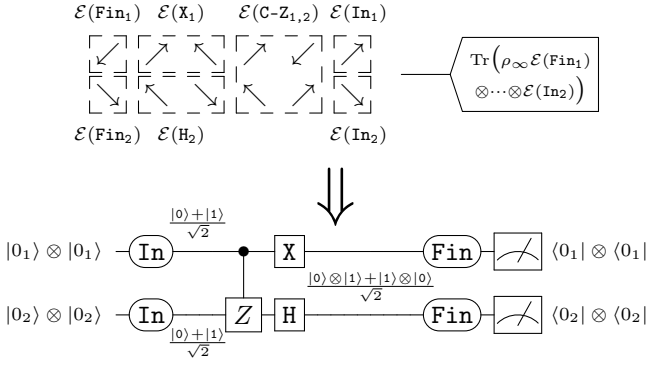

Even though the auxiliary space of a single chain in \eqref{NESS1} is four-dimensional, matrices \eqref{AA} can only span the full algebra of a two-dimensional subspace. Thus, we use \eqref{prod.seq} to encode the evolution of a single qubit in the auxiliary space of a single spin chain. In order to simulate a many-qubit circuit in the auxiliary spaces, the Hilbert space of each logical qubit is encoded in the auxiliary space of each spin chain: define

\begin{equation}
|\texttt{0}_\texttt{k}\rangle\equiv|0_k\rangle\otimes|1_k\rangle, \qquad |\texttt{1}_\texttt{k}\rangle\equiv|1_k\rangle\otimes|0_k\rangle,
\end{equation}
and denote by

\begin{equation}
\mathbbm{H}_\texttt{k}=\textnormal{span}\{|\texttt{0}_\texttt{k}\rangle,|\texttt{1}_\texttt{k}\rangle\}
\end{equation}
the $k$-th logical qubit Hilbert space. We use the typewriter font
to distinguish states and operators of the auxiliary space that
are relevant for the encoding of quantum circuits there. We prove that a quantum circuit, represented as a sequence of elementary operations $\prod_{\texttt{j}}\texttt{G}_{\texttt{j}}$ each acting on a small number of qubits, can be mapped in the auxiliary spaces, and its transition amplitudes can be computed via the expectation value of a tensor product of operators. Each of these operators $\mathcal{E}\left(\texttt{G}_\texttt{j}\right)$ encodes an elementary gate $\texttt{G}_\texttt{j}$, and acts on a small subset of physical spins, denoted by $\xi(\texttt{j})$, which belong to the chains representing the logical qubits transformed by the gate, as will be shown later:

\begin{equation}
\textnormal{Tr}\left(\rho_\infty\prod_{\xi(\texttt{j})}\mathcal{E}\left(\texttt{G}_\texttt{j}\right)
\right)=\frac{\big(\langle0|\langle0|\big)^{\otimes n}\left(\prod_\texttt{j}\texttt{G}_\texttt{j}\right)\big(|0\rangle|0\rangle\big)^{\otimes n}}{\prod_{k=1}^n\textnormal{Tr}\left(S_k S_k^\dag\right)}.\!\! \label{loc.seq}
\end{equation}
We will refer to the operations $\texttt{G}_\texttt{j}$ on the auxiliary spaces as \emph{encodings}, and to the corresponding operators $\mathcal{E}\left(\texttt{G}_\texttt{j}\right)$ on the physical spins as \emph{encoders}, see Fig. \ref{ex.circ}.

We want to stress that the mapping of quantum circuits to the auxiliary space relies upon the MPO structure of the NESS \eqref{NESS1}, which is a correspondence between Pauli matrices of physical spins and matrices $\mathbb{A}_{s}^{(k)}$ on the auxiliary space. The NESS \eqref{NESS1} has many matrix elements in the auxiliary space, like the right-hand-side of \eqref{prod.seq} for a single spin chain or of \eqref{loc.seq} for many spin chains, each one is a coefficient of a suitable operator basis expansion. These matrix elements from the auxiliary space can be extracted by means of expectation values, like the left-hand side of \eqref{loc.seq}. This piece of information is enough to estimate the complexity of measuring products of operators, each acting on a small number (at most four in the following discussion) of physical spins, thanks to the above mapping of quantum circuits and results from quantum computational complexity.

Since the amplitude on the right-hand-side of \eqref{loc.seq} is a complex number, the operator $\prod_\xi\mathcal{E}\left(\texttt{G}_\texttt{j}\right)$ is not necessarily Hermitian, thus not an observable. One can however choose normal matrices as encoders $\mathcal{E}\left(\texttt{G}_\texttt{j}\right)$, whose spectral decompositions are

\begin{equation}
\mathcal{E}\left(\texttt{G}_\texttt{j}\right)=\sum_eg(e_\xi)|e_\xi\rangle\langle e_\xi|
\end{equation}
with orthonormal eigenvectors $\{|e_\xi\rangle\}_e$ and complex eigenvalues $\{g(e_\xi)\}_e$. The procedure to get the amplitude on the right-hand-side of \eqref{loc.seq} from $\rho_\infty$ is the following.

\begin{enumerate}
\item[\emph{i)}] Perform a projective measurement onto the states $\{|e_\xi\rangle\}_e$ for each subset $\xi$ \footnote{It might happen that the encoder $\mathcal{E}(\texttt{G}_\texttt{j})$ has degenerate eigenvalues. In this case, it is not necessary to distinguish the corresponding eigenvectors during the measurement.}. We consider pairwise disjoint subsets $\{\xi\}$, so that the above measurements can be performed simultaneously.
\item[\emph{ii)}] Sample $M$ outcomes of these measurements, corresponding to sequences $|\{e_{\xi_1}^{(m)},e_{\xi_2}^{(m)},\dots\}\rangle$ for $m=1,\dots,M$, and compute
\begin{equation}
\Gamma=\frac{1}{M}\sum_{m=1}^M\prod_\xi g(e^{(m)}_\xi).
\end{equation}
\item[\emph{iii)}] The probability to obtain each of the $M$ outcomes is
\begin{equation} \label{prob}
p(\{e_\xi^{(m)}\}_\xi)=\langle\{e_\xi^{(m)}\}_\xi|\rho_\infty|\{e_\xi^{(m)}\}_\xi\rangle.
\end{equation}
We can see $\prod_\xi g(e^{(m)}_\xi)$ as a random variable sampled from the probability distribution $p(\{e_\xi^{(m)}\}_\xi)$, thus $\Gamma$ is a complex random variable with average
\begin{equation} \label{stat}
\mathbb{E}_p(\Gamma)=\frac{1}{M}\sum_{m=1}^M\mathbb{E}_p\Bigg(\prod_\xi g(e^{(m)}_\xi)\Bigg)=\textnormal{Tr}\Bigg[\rho_\infty\prod_{\xi(j)}\mathcal{E}\left(\texttt{G}_\texttt{j}\right)\Bigg].
\end{equation}
\end{enumerate}

The procedure \emph{i)-iii)} allows us to extract matrix elements from the auxiliary space, estimating suitable expectation values, as in equation \eqref{loc.seq}. This in turn requires the measurement of an operator, which, as any measurement, perturbs the state that is no longer a NESS after the above scheme.

The number of physical spins is $nL$, where $L$ is proportional to the number of elementary gates on the logical space $\bigotimes_{k=1}^n\mathbbm{H}_\texttt{k}$, because their encoders use only fixed numbers of physical spins.

In the following, we describe the encodings which fulfil the above \textit{requirements, i.e. encoders employing a small number of physical spins and to be normal matrices}.

\subsection{Encoder rules} \label{encoder.rules}

In the following subsections, we will develop encodings of specific and relevant gates as polynomials of matrices $\mathbb{A}^{(k)}_s$ in the auxiliary space of one or more chains. We now give a general recipe to derive the encodes $\mathcal{E}(\texttt{G})$ from any gate $\texttt{G}$. We start with a gate acting on a single logical qubit, thus using the auxiliary space of a single chain (the $k$-th one), and described by a product of matrices $\mathbb{A}^{(k)}_s$:

\begin{equation} \label{monomialA}
\texttt{G}^{(\texttt{k})}=\prod_{x=0}^X\mathbb{A}^{(k)}_{s_x}.
\end{equation}
The corresponding encoder is

\begin{equation} \label{monomialPauli}
\mathcal{E}(\texttt{G}^{(\texttt{k})})=\prod_{x=0}^X\big(\sigma^{s_x}_{k,j+x}\big)^\dag,
\end{equation}
where the product of matrices $\mathbb{A}^{(k)}_s$ is replaced by the product of the corresponding Pauli matrices in \eqref{NESS1}, and $j$ is the position where the encoder starts in the $k$-th chain. Since a general polynomial is a linear combination of monomials, the general single qubit gate

\begin{equation} \label{polynomialA}
\texttt{G}^{(\texttt{k})}=\sum_\texttt{y}\alpha_\texttt{y}\texttt{G}_\texttt{y}^{(\texttt{k})}
\end{equation}
with $\texttt{G}_\texttt{y}^{(\texttt{k})}$ being monomials \eqref{monomialA}, has encoder

\begin{equation}
\mathcal{E}(\texttt{G}^{(\texttt{k})})=\sum_\texttt{y}\alpha_\texttt{y}\mathcal{E}(\texttt{G}_\texttt{y}^{(\texttt{k})}).
\end{equation}

A gate acting on several logical qubits is encoded in the auxiliary spaces of the corresponding chains, each described by a NESS. The encoder of a gate that independently act on several qubits, e.g.

\begin{equation} \label{indep.qubit.gate}
\texttt{G}=\bigotimes_{\texttt{k}}\texttt{G}^{(\texttt{k})}
\end{equation}
where $\texttt{G}^{(\texttt{k})}$ is a general single qubit gate acting on the $k$-th logical qubit like in \eqref{polynomialA}, is

\begin{equation}
\mathcal{E}(\texttt{G})=\prod_{\texttt{k}}\mathcal{E}(\texttt{G}^{(\texttt{k})}).
\end{equation}
The latter product now pertains physical qubits belonging to the different chains that represent the logical qubits transformed by the gate. Finally, the encoder of a general linear combination

\begin{equation}
\texttt{G}=\sum_\texttt{k}\beta_\texttt{k}\bigotimes_\texttt{x}
\texttt{G}_\texttt{x}^{(\texttt{k})},
\end{equation}
with $\texttt{G}_\texttt{x}^{(\texttt{k})}$ being gates like \eqref{indep.qubit.gate}, is

\begin{equation}
\mathcal{E}(\texttt{G})=\sum_k\alpha_k\prod_\texttt{x}\mathcal{E}
\big(\texttt{G}_\texttt{x}^{(\texttt{k})}\big).
\end{equation}

The above discussion implies that the map $\mathcal{E}$ is linear, and satisfies the multiplicative properties $\mathcal{E}(G_1G_2)=\mathcal{E}(G_1)\mathcal{E}(G_2)$, i.e. products of operators on two horizontally contiguous groups of spins, and $\mathcal{E}(G_1\otimes G_2)=\mathcal{E}(G_1)\mathcal{E}(G_2)$, i.e. products of operators on two disjoint groups of chains.

\subsection{Initialization}

Notice that the amplitude resulting from the estimation \eqref{loc.seq} starts from and ends in the state

\begin{equation} \label{in.fin.state}
\bigotimes_{k=1}^n|0_k\rangle\otimes|0_k\rangle\notin\bigotimes_{k=1}^n\mathbbm{H}_\texttt{k}
\end{equation}
which is not in the logical computational space $\bigotimes_{k=1}^n\mathbbm{H}_\texttt{k}$. Thus, the first step of the computation in the auxiliary space must be a transition from $|0_k\rangle\otimes|0_k\rangle$ into a state

\begin{equation} \label{1qubit.state}
|\psi_\texttt{k}\rangle=a_k|\texttt{0}_\texttt{k}\rangle+b_k|\texttt{1}_\texttt{k}\rangle\in\mathbbm{H}_\texttt{k}
\end{equation}
for each chain. This transition corresponds to the encoding

\begin{equation}
\texttt{In}_{\texttt{k}}=2\big(a_k\mathbb{A}_+^{(k)}+b_k\mathbb{A}_-^{(k)}\big)\big|_{|0_k\rangle\otimes|0_k\rangle}
\end{equation}
with encoder

\begin{equation}
\mathcal{E}(\texttt{In}_{\texttt{k}})=2\big(a_k\sigma_{k,1}^-+b_k\sigma_{k,1}^+\big)
\end{equation}
which is normal if and only if $|a_k|=|b_k|$.

\subsection{Universal set of unitary gates}

Encodings of single qubit Pauli matrices on the logical space $\bigotimes_{k=1}^n\mathbbm{H}_\texttt{k}$ are

\begin{eqnarray}
\texttt{X}_{\texttt{k}} & = & \frac{32i}{\lambda^3+16\lambda}\left(\mathbb{A}_-^{(k)}\mathbb{A}_-^{(k)}-\mathbb{A}_+^{(k)}\mathbb{A}_+^{(k)}\right)\Big|_{\mathbbm{H}_\texttt{k}}, \nonumber \\
\texttt{Y}_{\texttt{k}} & = & -\frac{32}{\lambda^3+16\lambda}\left(\mathbb{A}_-^{(k)}\mathbb{A}_-^{(k)}+\mathbb{A}_+^{(k)}\mathbb{A}_+^{(k)}\right)\Big|_{\mathbbm{H}_\texttt{k}}, \nonumber \\
\texttt{Z}_{\texttt{k}} & = & \frac{32i}{\lambda^3+16\lambda}\left(\mathbb{A}_+^{(k)}\mathbb{A}_-^{(k)}+\mathbb{A}_-^{(k)}\mathbb{A}_+^{(k)}\right)\Big|_{\mathbbm{H}_\texttt{k}}, \nonumber \\
\texttt{1}_{\texttt{k}} & = & \frac{32i}{\lambda^3+16\lambda}\left(\mathbb{A}_+^{(k)}\mathbb{A}_-^{(k)}-\mathbb{A}_-^{(k)}\mathbb{A}_+^{(k)}\right)\Big|_{\mathbbm{H}_\texttt{k}}. \label{Pauli}
\end{eqnarray}

An example of two-qubit gate is the controlled-Z gate which has the following encoding

\begin{eqnarray} \label{CZ}
\texttt{C-Z}_{\texttt{k},\texttt{l}} & = & |\texttt{0}_\texttt{k}\rangle\langle\texttt{0}_\texttt{k}|\otimes\texttt{1}_{\texttt{l}}+|\texttt{1}_\texttt{k}\rangle\langle\texttt{1}_\texttt{k}|\otimes\texttt{Z}_{\texttt{l}}= \nonumber \\
& = & \frac{1}{2}\left(\texttt{1}_{\texttt{k}}+\texttt{Z}_{\texttt{k}}\right)\otimes\texttt{1}_{\texttt{l}}+\frac{1}{2}\left(\texttt{1}_{\texttt{k}}-\texttt{Z}_{\texttt{k}}\right) \otimes\texttt{Z}_{\texttt{l}}
\end{eqnarray}

Single qubit unitaries

\begin{equation}
e^{i\theta\texttt{G}_k}=\cos(\theta)\texttt{1}_k+i\sin(\theta)\texttt{G}_k
\end{equation}
with $\texttt{G}\in\{\texttt{X},\texttt{Y},\texttt{Z}\}$ and $\theta\in\mathbbm{R}$, and the controlled-Z gate \eqref{CZ}, form a universal set \cite{Barenco1995} in the sense that they can be multiplied to generate any many-qubit unitary \footnote{The universal set proved in \cite{Barenco1995} is made of single qubit unitaries and the controlled-NOT gate, Nevertheless, we can substitute the controlled-NOT gate with the controlled-Z gate since $\texttt{C-Z}_{\texttt{k},\texttt{l}}=\texttt{1}_{\texttt{k}}\otimes e^{i\frac{\pi}{4}\texttt{Y}_{\texttt{l}}}\cdot\texttt{C-NOT}_{\texttt{k},\texttt{l}}\cdot\texttt{1}_{\texttt{k}}\otimes e^{-i\frac{\pi}{4}\texttt{Y}_{\texttt{l}}}$. See also the discussion on the normality of encoders in appendix \ref{app-encoders}.}. The above universal set has normal encoders, see section \ref{encoder.rules} and appendix \ref{app-encoders}, and thus form the elementary operations to implement a quantum algorithm.

\subsection{Measurements and postselection}

As measurements, that can be applied during the algorithm, we consider projective measurements without loss of generality \cite{NielsenChuang}. The projector $|\psi_\texttt{k}\rangle\langle\psi_\texttt{k}|$, with $|\psi_\texttt{k}\rangle$ being a general state of a single logical qubit as in \eqref{1qubit.state}, has encoding

\begin{eqnarray}
|\psi_\texttt{k}\rangle\langle\psi_\texttt{k}| & = & \frac{|a_k|^2}{2}\left(\texttt{1}_{\texttt{k}}+\texttt{Z}_{\texttt{k}}\right)+\frac{a_k\overline{b}_k}{2}\left(\texttt{X}_{\texttt{k}}+i \, \texttt{Y}_{\texttt{k}}\right) \nonumber \\
&& +\frac{\overline{a}_kb_k}{2}\left(\texttt{X}_{\texttt{k}}-i \, \texttt{Y}_{\texttt{k}}\right)+\frac{|b_k|^2}{2}\left(\texttt{1}_{\texttt{k}}-\texttt{Z}_{\texttt{k}}\right), \label{proj}
\end{eqnarray}
with normal encoder if and only if $|a_k|=|b_k|$, see appendix \ref{app-encoders}. The latter condition leaves enough freedom to implement projectors onto some complete orthonormal bases on each qubit. The other bases can be reduced to these by applying unitary operations.

The last measurement leading to the read-out of the result in a quantum circuit deserves a special mention. Indeed, the very last step of the computation in the auxiliary state is the projection onto the fixed state \eqref{in.fin.state}, because of the MPO structure in \eqref{NESS1}. In the last operations before this projection, any state can be rotated into \eqref{in.fin.state}. For instance, first rotate the desired state into a tensor product of states $|\psi_\texttt{k}\rangle$ in \eqref{1qubit.state} by means of unitary operations; then transform each state $|\psi_\texttt{k}\rangle$ into

\begin{equation}
|0_k\rangle\otimes|0_k\rangle+\frac{1}{2}|1_k\rangle\otimes|1_k\rangle
\end{equation}
via the operation

\begin{equation}
\texttt{Fin}_\textnormal{k}=\frac{4i}{\lambda}\big(a_k\mathbb{A}_-^{(k)}-b_k\mathbb{A}_+^{(k)}\big)\big|_{\mathbbm{H}_{\texttt{k}}}
\end{equation}
with encoder

\begin{equation}
\mathcal{E}(\texttt{Fin}_\textnormal{k})=\frac{4i}{\lambda}\big(a_k\,\sigma_{k,L}^+-b_k\,\sigma_{k,L}^-\big)
\end{equation}
which is normal if and only if $|a_k|=|b_k|$.

The above procedures however project onto a controlled state. This abtract ability, called \emph{postselection}, is not believed to be implemented with a quantum computer, and enourmously increases the power of realistic quantum computer \cite{Aaronson2005}. Nevertheless, measurements in quantum circuits lead to random projections onto the computational basis. This randomness cannot be simulated in our scheme, but we will discuss how to cope with this impossibility. Furthermore, postselection can also be implemented before the end of the computation by encoding a fixed projector \eqref{proj} in the auxiliary space.

\subsection{Uncomputing and error reduction} \label{uncomp.err.red}

Two important ingredients of standard quantum computation are \emph{uncomputing} and \emph{error reduction} \cite{Bennett1997}. The former is the ability to undo the computation after having backed up the computational result, the latter is the reduction of the error probability due to the probabilistic nature of quantum measurements.






As mentioned in the previous subsection, the read-out measurement results in a random projection but this randomness cannot be simulated in the auxiliary space since only fixed projections can be implemented. Moreover, the computational result of decision problems is encoded in a single logical qubit, the only one which is needed to be measured. Nevertheless, the MPO structure in \eqref{NESS1} allows us to extract only matrix elements \eqref{loc.seq} with the leftmost state of all the chains, thus of all the logical qubits, being \eqref{in.fin.state}. Uncomputing allows us to achieve a final state that is known except for the logical qubit encoding the computational result, and thus only two matrix elements, varying the state of this logical qubit in the computational basis, is required to find the computational result.

The uncomputing procedure for standard quantum computation without postselection can be implemented. Nevertheless, we aim to encode quantum computation with postselection in the auxiliary space, and standard uncomputing cannot be straightforwardly generalised in the presence of postselection. The reason is that the unitarity of quantum circuits is crucial there \cite{Bennett1997}, but quantum circuits with postselection are not in general unitary. We thus implement uncomputing with a non-unitary operation on the auxiliary space, resorting to a novel scheme suited for quantum computation with postselection.
First, represent a quantum circuit with postselection by a unitary $\texttt{U}$ followed by the postselection of the first logical qubit in the state $|\texttt{1}\rangle$: recall that the postselection can be performed at the end of the computation and on a single qubit without loss of generality \cite{Aaronson2005}. Denote by $\mathcal{L}$ the decision problem solved by $\texttt{U}$ followed by the above postselection, such that $\mathcal{L}(\texttt{x})\in\{\texttt{0},\texttt{1}\}$ is the correct answer, having specified the problem with the input $\texttt{x}$.

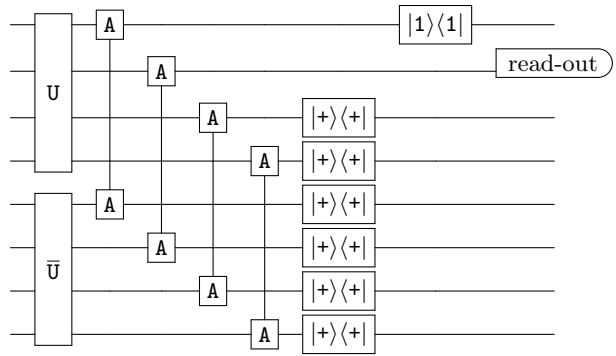
\begin{figure}[htbp]
\[
\Qcircuit @C=1em @R=0.2em {
& \multigate{3}{\texttt{U}} & \gate{\texttt{A}} & \qw & \qw & \qw & \qw & \gate{|\texttt{1}\rangle\langle\texttt{1}|} & \qw \\
& \ghost{\texttt{U}} & \qw \qwx & \gate{\texttt{A}} & \qw & \qw & \qw & \qw & \measureD{\textnormal{read-out}} \\
& \ghost{\texttt{U}} & \qw \qwx & \qw \qwx & \gate{\texttt{A}} & \qw & \gate{|\texttt{+}\rangle\langle\texttt{+}|} & \qw & \qw \\
& \ghost{\texttt{U}} & \qw \qwx & \qw \qwx & \qw \qwx & \gate{\texttt{A}} & \gate{|\texttt{+}\rangle\langle\texttt{+}|} & \qw & \qw \\
& \multigate{3}{\overline{\texttt{U}}} & \gate{\texttt{A}} \qwx & \qw \qwx & \qw \qwx & \qw \qwx & \gate{|\texttt{+}\rangle\langle\texttt{+}|} & \qw & \qw \\
& \ghost{\overline{\texttt{U}}} & \qw & \gate{\texttt{A}} \qwx & \qw \qwx & \qw \qwx & \gate{|\texttt{+}\rangle\langle\texttt{+}|} & \qw & \qw \\
& \ghost{\overline{\texttt{U}}} & \qw & \qw & \gate{\texttt{A}} \qwx & \qw \qwx  & \gate{|\texttt{+}\rangle\langle\texttt{+}|} & \qw & \qw \\
& \ghost{\overline{\texttt{U}}} & \qw & \qw & \qw & \gate{\texttt{A}} \qwx & \gate{|\texttt{+}\rangle\langle\texttt{+}|} & \qw & \qw
}
\]
\caption{Uncomputing procedure for quantum computation with postselection.}
\label{ampl-fig}
\end{figure}

The uncomputing procedure shown in figure \ref{ampl-fig} is the following.

\begin{enumerate}
\item[\emph{i)}] Encode two circuits, $\texttt{U}$ and its complex conjugate $\overline{\texttt{U}}$, in parallel, which produce the state
\begin{equation} \label{final.uncomp}
\sum_{\texttt{y}\in\{\texttt{0},\texttt{1}\}^n}c_\texttt{y}|\texttt{y}\rangle\otimes\sum_{\texttt{y}\in\{\texttt{0},\texttt{1}\}^n}\bar c_\texttt{y}|\texttt{y}\rangle.
\end{equation}
\item[\emph{ii)}] Encode the following operator jointly on the $k$-th qubit of both circuits $\forall\,\texttt{k}$
\begin{align}
\texttt{A}_{\texttt{k},\texttt{n+k}}= & |\texttt{0}_\texttt{k}\rangle\langle\texttt{0}_\texttt{k}|\otimes|\texttt{0}_\texttt{n+k}\rangle\langle\texttt{0}_\texttt{n+k}|+|\texttt{1}_\texttt{k}\rangle\langle\texttt{1}_\texttt{k}|\otimes|\texttt{1}_\texttt{n+k}\rangle\langle\texttt{1}_\texttt{n+k}| \nonumber \\
= & \frac{\texttt{1}_\texttt{k}+\texttt{Z}_\texttt{k}}{2}\otimes\frac{\texttt{1}_\texttt{n+k}+\texttt{Z}_\texttt{n+k}}{2}+\frac{\texttt{1}_\texttt{k}-\texttt{Z}_\texttt{k}}{2}\otimes\frac{\texttt{1}_\texttt{n+k}-\texttt{Z}_\texttt{n+k}}{2} \label{Aop}
\end{align}
whose encoder is derived as described in section \ref{encoder.rules}, see also appendix \ref{app-encoders}, and produces the state
\begin{equation}
\sum_{\texttt{y}\in\{\texttt{0},\texttt{1}\}^n}|c_\texttt{y}|^2|\texttt{y}\rangle\otimes|\texttt{y}\rangle.
\end{equation}
Note that the latter state is not normalised, but this is not a problem because it lies on the auxiliary, virtual, space, not on the physical space.
\item[\emph{iii)}] We apply the projector $|\texttt{+}\rangle\langle\texttt{+}|$ to all logical qubits except the qubit of the first circuit to be postselected and the qubit of the first circuit that records the result. The operation $\texttt{A}$ is required to create a state with only positive coefficients in the computational basis, such that the projection onto $|\texttt{+}\rangle$ never has vanishing amplitude. Then, after the postselection, and denoting by $\neg\mathcal{L}(\texttt{x})$ the negation of $\mathcal{L}(\texttt{x})$ and by $\circ$ the concatenation of bit strings, we obtain the state
\begin{eqnarray}
&& \frac{1}{2^{n-1}} \, |\texttt{1}\rangle\otimes\sum_{\texttt{y}\in\{\texttt{0},\texttt{1}\}^{n-2}}\Big(|c_{\texttt{1}\circ\mathcal{L}(\texttt{x})\circ\texttt{y}}|^2\big|\mathcal{L}(\texttt{x})\big\rangle \nonumber \\
&& +|c_{\texttt{1}\circ\neg\mathcal{L}(\texttt{x})\circ\texttt{y}}|^2\big|\neg\mathcal{L}(\texttt{x})\big\rangle\Big)\otimes|\texttt{+}\rangle^{\otimes(2n-2)}. \label{uncomp.state}
\end{eqnarray}
\end{enumerate}

For standard quantum computation without postselection, error reduction is implemented by many repetitions of the quantum circuit and by applying majority vote to the several outcomes in order to decide the solution \cite{Bennett1997,Bernstein1997,AroraBarak}. The Chernoff bound implies that if the single-circuit success probability is at least $1/2+1/\textnormal{poly}(n)$, the success probability resulting from the majority vote with polynomially many (in $n$) repetitions is exponentially (in $n$) close to one. The same error reduction procedure can be applied for quantum circuits with postselection. Furthermore, note that the success and failure probabilities, $p$ and $1-p$ respectively, of the original circuit with postselection are

\begin{align}
p & =\frac{\displaystyle \sum_{\texttt{y}\in\{\texttt{0},\texttt{1}\}^{n-2}}|c_{\texttt{1}\circ\mathcal{L}(\texttt{x})\circ\texttt{y}}|^2}{\displaystyle \sum_{\texttt{y}\in\{\texttt{0},\texttt{1}\}^{n-1}}|c_{\texttt{1}\circ\texttt{y}}|^2}, \label{succ.prob} \\
1-p & = \frac{\displaystyle \sum_{\texttt{y}\in\{\texttt{0},\texttt{1}\}^{n-2}}|c_{\texttt{1}\circ\neg\mathcal{L}(\texttt{x})\circ\texttt{y}}|^2}{\displaystyle \sum_{\texttt{y}\in\{\texttt{0},\texttt{1}\}^{n-1}}|c_{\texttt{1}\circ\texttt{y}}|^2}, \label{err.prob}
\end{align}
and the success and failure probabilities, $p'$ and $1-p'$ respectively, from the state \eqref{uncomp.state}, after normalisation, are

\begin{align}
p' & =\frac{\displaystyle \sum_{\texttt{y}\in\{\texttt{0},\texttt{1}\}^{n-2}}|c_{\texttt{1}\circ\mathcal{L}(\texttt{x})\circ\texttt{y}}|^4}{\displaystyle \sum_{\texttt{y}\in\{\texttt{0},\texttt{1}\}^{n-1}}|c_{\texttt{1}\circ\texttt{y}}|^4}, \label{succ.prob1} \\
1-p' & = \frac{\displaystyle \sum_{\texttt{y}\in\{\texttt{0},\texttt{1}\}^{n-2}}|c_{\texttt{1}\circ\neg\mathcal{L}(\texttt{x})\circ\texttt{y}}|^4}{\displaystyle \sum_{\texttt{y}\in\{\texttt{0},\texttt{1}\}^{n-1}}|c_{\texttt{1}\circ\texttt{y}}|^4}. \label{err.prob1}
\end{align}
From the comparison between $p$ and $p'$, another consequence of the operation $\texttt{A}$ \eqref{Aop} is an exponential increase of the success probability. Thus, polynomially many (in $n$) applications of the operation $\texttt{A}$ \eqref{Aop} provide an alternative error reduction procedure for quantum computation with postselection.

\subsection{Computational complexity}

Before discussing the computational complexity of the above quantum circuit encoding, we recall some basic computational classes. \textsf{P} is the class of decision problems, i.e. with yes/no answers, solved by a polynomial-time deterministic Turing machine, \textsf{NP} is the class of decision problems solved by at least one path of a polynomial-time non-deterministic (i.e. with multiple transitions) Turing machine, \textsf{\#{}P} consists of functions counting the number of accepting paths of \textsf{NP} problems, \textsf{PP} is the class of decision problems solved with success probability larger than $\frac{1}{2}$ by a polynomial-time probabilistic Turing machine \cite{AroraBarak,DuKo}. The power of quantum computers is captured by the computational class \textsf{BQP}, defined by decision problems solved by a polynomial-time quantum Turing machine with success probability at least $\frac{2}{3}$ \cite{Bennett1997,Bernstein1997,AroraBarak}. Turing machines in the definition of certain complexity classes, e.g. \textsf{P} and \textsf{BQP}, can be complemented with calls to oracles able to solve hard problems, e.g. $\mathcal{O}$, or entire classes such as \textsf{PP} or \textsf{\#{}P}. These generalisations give rise to new classes, like $\textsf{BQP}^\mathcal{O}$ or $\textsf{P}^\textsf{PP}$, where the exponents represent the power of the oracles \cite{AroraBarak,DuKo}. An interesting equality between classes of problem solved by oracle machines is $\textsf{P}^\textsf{PP}=\textsf{P}^\textsf{\#{}P}$ \cite{DuKo}, suggesting that the capabilities to solve problems in \textsf{PP} and to solve those in $\textsf{\#{}P}$ are the same.

The threshold probability $\frac{2}{3}$ in the definition of \textsf{BQP} is arbitrary, and can become exponentially (in $n$) close to $1$ with polynomially many (in $n$) repetitions and majority vote, technically via the Chernoff bound \cite{Bennett1997,Bernstein1997,AroraBarak}. Using this \emph{error reduction} and \emph{uncomputing}, i.e. the ability to undo the computation after having stored the computational result, any quantum algorithm, solving a problem in \textsf{BQP}, can be efficiently transformed such that the final state is a superposition with the squared amplitude of the state $|\mathcal{L}(\texttt{x})\rangle\otimes|\texttt{x}\rangle$ greater than $1-\varepsilon$, $\texttt{x}\in\{0,1\}^{n-1}$ being the problem input of $n-1$ bits, $\mathcal{L}(\texttt{x})$ being $1$ ($0$) if the problem solution is yes (no), and $\varepsilon$ being exponentially small in $n$ \cite{Bennett1997}. As discussed in section \ref{uncomp.err.red}, uncomputing aims to circumvent limitations imposed by the MPO structure in \eqref{NESS1}, namely that the projection onto a fixed state is implemented as the last operation in the auxiliary space instead of a proper read-out measurement. We thus encode the aforementioned transformed algorithm in the auxiliary space of NESS, finally project onto the state $|\mathcal{L}(\texttt{x})=\texttt{1}\rangle\otimes|\texttt{x}\rangle$, and find the solution depending on whether the squared modulus of the transition amplitude, measured through the expectation of the corresponding local operator \eqref{loc.seq}, is greater than $1-\varepsilon$.

We have seen that measuring local operators \eqref{loc.seq} of NESS provides computational capability comparable to that of quantum computers. We now show stronger results. We can encode in the auxiliary space of NESS both heralded projections \eqref{proj}, simulating postselection of measurement outcomes with unit probability, as well as general linear invertible operations, using linear combinations of single qubit Pauli matrices \eqref{Pauli} and of their tensor products. The additional power provided to quantum Turing machine by both types of operations is equivalent to solve any problem in \textsf{PP} in polynomial time and with bounded error probability (in analogy with the definition of \textsf{BQP}) \cite{Aaronson2005}.

We can encode in the auxiliary space the quantum circuit with postseletion followed by the uncomputing procedure described in section \ref{uncomp.err.red}, and measure, via \eqref{loc.seq}, the transition amplitudes with the final states, $|\texttt{1}\rangle\otimes|\mathcal{L}(\texttt{x})=\texttt{0}\rangle\otimes|\texttt{+}\rangle^{\otimes 2n-2}$ and $|\texttt{1}\rangle\otimes|\mathcal{L}(\texttt{x})=\texttt{1}\rangle\otimes|\texttt{+}\rangle^{\otimes 2n-2}$ respectively, to be transformed into $|\texttt{+}\rangle^{\otimes 2n}$ in the very last step of the encoding \footnote{Note that the number of logical qubits is doubled compared to the notation of \eqref{loc.seq}, due to the uncomputing procedure.}. The ratio between these transition amplitudes identifies the correct answer of any \textsf{PP} language, i.e. whether $\mathcal{L}(\texttt{x})=\texttt{0}$ or $\mathcal{L}(\texttt{x})=\texttt{1}$. Note that the first logical qubit, in the state $|\texttt{1}\rangle$, can be transformed into $|\texttt{+}\rangle$ but it is not necessary. Indeed, the transition amplitude in the auxiliary space, as in equation \eqref{loc.seq}, ends with $\langle\texttt{+}|$ for the auxiliary space of each chain, encoding each logical qubit. Without transforming the final state of the first logical qubit $|\texttt{1}\rangle$ into $|\texttt{+}\rangle$, we simply get a constant factor $1/\sqrt{2}$ that can be reabsorbed in the proportionality constant of equation \eqref{loc.seq}.

Now, we show that the estimation of \eqref{loc.seq} also allows us to compute \textsf{\#{}P} functions which count the number of accepting paths of an \textsf{NP} problem. As an example of \textsf{NP-complete} problem, consider the Boolean satisfiability problem, i.e. the satisfiability of a Boolean function $f(x):\{0,1\}^n\to\{0,1\}$. The corresponding \textsf{\#{}P-complete} problem is the computation of the number of satisfying assignments $s(f)=\big|\{x\in\{0,1\}^n : f(x)=1\}\big|$. In order to compute $s(f)$, first encode the state

\begin{equation}
\frac{1}{2^{\frac{n}{2}}}\sum_{\texttt{x}\in\{\texttt{0},\texttt{1}\}^n}|\texttt{x}\rangle\otimes|\texttt{0}\rangle
\end{equation}
in the auxiliary space of spin chains, and then compute the Boolean function with standard logical gates, producing the state

\begin{equation}
\frac{1}{2^{\frac{n}{2}}}\sum_{\texttt{x}\in\{\texttt{0},\texttt{1}\}^n}|\texttt{x}\rangle\otimes|f(\texttt{x})\rangle.
\end{equation}
Applying the steps $i)$ and $ii)$ of Section \ref{uncomp.err.red}, we obtain the state

\begin{equation} \label{state.count}
\frac{1}{2^{2n+\frac{1}{2}}}|\texttt{+}\rangle^{2n+1}\otimes\big(s(f)|1\rangle+\big(2^{n}-s(f)\big)|0\rangle\big).
\end{equation}
The state \eqref{state.count}, up to normalisation, can also be obtained using postselection \citep{Aaronson2005}, following \cite{Abrams1998}, so with operations \eqref{proj}. Therefore, measuring transition amplitudes by means of \eqref{loc.seq}, we compute $s(f)$.

Thus, measuring local operators \eqref{loc.seq} allows us to solve problems in \textsf{PP} and \textsf{\#{}P}. We could also use expectations of local operations \eqref{loc.seq} as oracles for deterministic Turing machines, as models of classical computers, and thus solve problems in $\textsf{P}^\textsf{PP}=\textsf{P}^\textsf{\#{}P}$. Since $\textsf{P}^\textsf{PP}=\textsf{P}^\textsf{\#{}P}\supseteq\textsf{PP}^\textsf{PH}\supseteq\textsf{PH}$, where \textsf{PH} (polynomial hierarchy) is the union of a, conjectured infinite, hierarchy of classes that generalise \textsf{P} and \textsf{NP} \cite{AroraBarak,DuKo}, the measure of expectations of local operations of NESS \eqref{loc.seq} would help to solve extremely hard problems.

\section{Conclusions}

We study the NESS of the XX spin chains with boundary noise from the computational complexity point of view. We base our study on the MPO structure of the NESS, and analyse how to efficiently encode the circuit model of a quantum Turing machine in the auxiliary space of NESS. We use $n$ independent chains, one for each qubit in the computation, and the length $L$ of the chains is proportional to the number of elementary gates, thus to the time complexity, of the encoded algorithm. The expectation value of certain operators, being local in the sense that independently act on small and disjoint sets of spins, is proportional to transition amplitudes of the above encoded algorithms.

Using polynomially many (in the number $n$ of logical qubits) spins and with the ability of estimating expectations of local operators, we can find solutions of all problems solvable by quantum computers and even much harder problems. If these problems are not believed to be solved efficiently by physical computers \cite{Aaronson2005-2}, our result implies that the required estimations in NESS are experimentally and operationally very difficult. Possible bottlenecks, running in superpolynomial time, are the estimation of expectations of local operators \eqref{loc.seq} itself and the relaxation time to NESS. The latter is ruled out because the spectral gap of the Liouvillian, namely the smallest non-vanishing real part of its eigenvalues, goes to zero as one divided by a polynomial of the chain length \cite{Prosen2008}. Thus, superpolynomially many measurement outcomes $M$ can be required for the accurate estimation of \eqref{stat}.

The need of superpolynomially many measurement outcomes $M$ is plausible but not evident a priori. Indeed, it is difficult to exclude that the expectation values \eqref{stat} can be measured with high precision or that the probability distribution \eqref{prob} can be compressed in order to efficiently estimate \eqref{loc.seq}. Therefore, we used computational complexity arguments for a deeper analysis, establishing which computational problems the measure of expectation values of local operators \eqref{loc.seq} allows us to solve, i.e. \textsf{PP} languages and \textsf{\#{}P} functions. This result provides a characterization of the hardness of measuring local operators \eqref{loc.seq} in computational complexity terms, which is more precise than the simple realisation of the need of superpolynomially many measurement outcomes $M$.

Furthermore, the present approach is the first, at the best of our knowledge, study of the computational complexity of NESS, which is also a non-equilibrium extension of equilibrium results on the complexity of ground states \cite{Kempe2006,Oliveira2008,Schuch2009,Whitfield2013,Childs2016} as well as a mixed state generalisation of the computational complexity study of pure states like PEPS \cite{Verstraete2006,Schuch2007}. We remind that the NESS is not entangled for small coupling $\lambda$ \cite{Prosen2011a,Marzolino2014}, and has no nearest neighbour spin entanglement for a wide range of parameters \cite{Znidaric2012}. Nevertheless quantum NESS give rise to interesting phenomena, such as non-equilibrium phase transitions, enhanced state distinguishability, and enhanced metrological performances \cite{Banchi2014,Marzolino2014}.

We only considered the XX model, $\Delta=0$, that has the minimal bond dimension of the MPO. However, for $\arccos\Delta=\frac{\pi l}{m}$ with $l,m$ coprime integers, each auxiliary space has dimension $m+1$ and is spanned by $\{|0_k\rangle,|1_k\rangle,|2_k\rangle,\dots,|m_k\rangle\}$, while in all the other cases the auxiliary spaces have dimension $\lfloor\frac{L}{2}\rfloor+1$. There can be denser encodings of more than one logical qubit in a single spin chain with higher dimensional auxiliary spaces, in order to minimise the number $n$ of spin chains. Since matrices $A_{s}^{(k)}$ allow only for transitions between contiguous levels, such denser encodings may increase the number $L$ of physical spins in each chain, inducing a trade-off between the minimisation of $L$ and that of $n$.

\appendix

\section{Encoders} \label{app-encoders}

In this section, we write explicitly the encoders of some logical gates, that are constructed according to the encoder rules described in section \ref{encoder.rules}. All the gates from the logical computational space into itself can be written as linear combinations of tensor products of the following four operators

\begin{eqnarray} \label{encoding}
|\texttt{0}_\texttt{k}\rangle\langle\texttt{0}_\texttt{k}| & = & \frac{32i}{\lambda^3+16\lambda}\mathbb{A}_+^{(k)}\mathbb{A}_-^{(k)}\Big|_{\mathbbm{H}_\texttt{k}}, \nonumber \\
|\texttt{0}_\texttt{k}\rangle\langle\texttt{1}_\texttt{k}| & = & -\frac{32i}{\lambda^3+16\lambda}\mathbb{A}_+^{(k)}\mathbb{A}_+^{(k)}\Big|_{\mathbbm{H}_\texttt{k}}, \nonumber \\
|\texttt{1}_\texttt{k}\rangle\langle\texttt{0}_\texttt{k}| & = & \frac{32i}{\lambda^3+16\lambda}\mathbb{A}_-^{(k)}\mathbb{A}_-^{(k)}\Big|_{\mathbbm{H}_\texttt{k}}, \nonumber \\
|\texttt{1}_\texttt{k}\rangle\langle\texttt{1}_\texttt{k}| & = & -\frac{32i}{\lambda^3+16\lambda}\mathbb{A}_-^{(k)}\mathbb{A}_+^{(k)}\Big|_{\mathbbm{H}_\texttt{k}}.
\end{eqnarray}
Their encoders require only two physical spins:

\begin{eqnarray} \label{encoders}
\mathcal{E}\left(|\texttt{0}_\texttt{k}\rangle\langle\texttt{0}_\texttt{k}|\right) & = & \frac{32i}{\lambda^3+16\lambda}
\sigma_{k,j}^-\sigma_{k,j+1}^+, \nonumber \\
\mathcal{E}\left(|\texttt{0}_\texttt{k}\rangle\langle\texttt{1}_\texttt{k}|\right) & = & -\frac{32i}{\lambda^3+16\lambda}
\sigma_{k,j}^-\sigma_{k,j+1}^-, \nonumber \\
\mathcal{E}\left(|\texttt{1}_\texttt{k}\rangle\langle\texttt{0}_\texttt{k}|\right) & = & \frac{32i}{\lambda^3+16\lambda}
\sigma_{k,j}^+\sigma_{k,j+1}^+, \nonumber \\
\mathcal{E}\left(|\texttt{1}_\texttt{k}\rangle\langle\texttt{1}_\texttt{k}|\right) & = & -\frac{32i}{\lambda^3+16\lambda}
\sigma_{k,j}^+\sigma_{k,j+1}^-.
\end{eqnarray}

The above encoders \eqref{encoders} and the linearity of the map $\mathcal{E}$ give the encoders of Pauli matrices \eqref{Pauli} on the logical qubits:

\begin{eqnarray}
\mathcal{E}\left(\texttt{X}_{\texttt{k}}\right) & = & \frac{32i}{\lambda^3+16\lambda}\left(\sigma_{k,j}^+\sigma_{k,j+1}^+
-\sigma_{k,j}^-\sigma_{k,j+1}^-\right), \nonumber \\
\mathcal{E}\left(\texttt{Y}_{\texttt{k}}\right) & = & -\frac{32}{\lambda^3+16\lambda}\left(\sigma_{k,j}^+\sigma_{k,j+1}^+
+\sigma_{k,j}^-\sigma_{k,j+1}^-\right), \nonumber \\
\mathcal{E}\left(\texttt{Z}_{\texttt{k}}\right) & = & \frac{32i}{\lambda^3+16\lambda}\left(\sigma_{k,j}^-\sigma_{k,j+1}^+
+\sigma_{k,j}^+\sigma_{k,j+1}^-\right), \nonumber \\
\mathcal{E}\left(\texttt{1}_{\texttt{k}}\right) & = & \frac{32i}{\lambda^3+16\lambda}\left(\sigma_{k,j}^-\sigma_{k,j+1}^+
-\sigma_{k,j}^+\sigma_{k,j+1}^-\right). \qquad
\end{eqnarray}
Using again the linearity of the map $\mathcal{E}$, we find the encoders of any single qubit unitary:

\begin{eqnarray}
\mathcal{E}\left(e^{i\theta\texttt{G}_k}\right) & = & \cos(\theta)\mathcal{E}\left(\texttt{1}_k\right)+
i\sin(\theta)\mathcal{E}\left(\texttt{G}_k\right), \quad \textnormal{with} \nonumber \\
e^{i\theta\texttt{G}_k} & = & \cos(\theta)\texttt{1}_k+
i\sin(\theta)\texttt{G}_k,
\end{eqnarray}
and $\texttt{G}\in\{\texttt{X},\texttt{Y},\texttt{Z}\}$.

The encoder of the control-Z gate \eqref{CZ} is
\begin{eqnarray}
&& \mathcal{E}\left(\texttt{C-Z}_{\texttt{k},\texttt{l}}\right)=\frac{1024}{(\lambda^3+16\lambda)^2}\left(\sigma_{k,j}^-\,\sigma_{k,j+1}^+\,\sigma_{l,j}^+
\,\sigma_{l,j+1}^- \right. \nonumber \\
&& -\sigma_{k,j}^-\,\sigma_{k,j+1}^+\,\sigma_{l,j}^-\,\sigma_{l,j+1}^+
-\sigma_{k,j}^+\,\sigma_{k,j+1}^-\,\sigma_{l,j}^-\,\sigma_{l,j+1}^+ \nonumber \\
&& \left. +\sigma_{k,j}^+\,\sigma_{k,j+1}^-\,\sigma_{l,j}^+\,\sigma_{l,j+1}^-\,\right). \label{CZ-enc}
\end{eqnarray}
All these encoders are normal matrices, as required in order to measure their expectation value.

The encoder of the projector $|\psi_\texttt{k}\rangle\langle\psi_\texttt{k}|$ \eqref{proj} with $|\psi_\texttt{k}\rangle=a_k|\texttt{0}_\texttt{k}\rangle+b_k|\texttt{1}_\texttt{k}\rangle\in\mathbbm{H}_\texttt{k}$ is

\begin{eqnarray}
\mathcal{E}\left(|\psi_\texttt{k}\rangle\langle\psi_\texttt{k}|\right) & = & \frac{32i}{\lambda^3+16\lambda}\left(|a_k|^2 \, \sigma_{k,j}^-\sigma_{k,j+1}^+ \right. \nonumber \\
\!\!\! && -a_k\overline{b}_k \, \sigma_{k,j}^-\sigma_{k,j+1}^-+\overline{a}_kb_k \, \sigma_{k,j}^+\sigma_{k,j+1}^+ \nonumber \\
&& \left. -|b_k|^2 \, \sigma_{k,j}^+\sigma_{k,j+1}^-\right),
\end{eqnarray}
which is normal if and only if $|a_k|=|b_k|$.

In order to see an example of non-normal encoder, use \eqref{encoding} for the encoding of the controlled-NOT gate

\begin{eqnarray}
\!\!\!\!\!\!\! && \texttt{C-NOT}_{\texttt{k},\texttt{l}}=|\texttt{0}_\texttt{k}\rangle\langle\texttt{0}_\texttt{k}|\otimes\texttt{1}_{\texttt{l}}+|\texttt{1}_\texttt{k}\rangle\langle\texttt{1}_\texttt{k}|\otimes\texttt{X}_{\texttt{l}}= \nonumber \\
\!\!\!\!\!\!\! && =\frac{1024}{(\lambda^3+16\lambda)^2}\left(\mathbb{A}_+^{(k)}\mathbb{A}_-^{(k)}\otimes\mathbb{A}_-^{(l)}\mathbb{A}_+^{(l)}-\mathbb{A}_+^{(k)}\mathbb{A}_-^{(k)}\otimes\mathbb{A}_+^{(l)}\mathbb{A}_-^{(l)} \right. \nonumber \\
\!\!\!\!\!\!\! && \left. -\mathbb{A}_-^{(k)}\mathbb{A}_+^{(k)}\otimes\mathbb{A}_+^{(l)}\mathbb{A}_+^{(l)}+\mathbb{A}_-^{(k)}\mathbb{A}_+^{(k)}\otimes\mathbb{A}_-^{(l)}\mathbb{A}_-^{(l)}\right)\Big|_{\mathbbm{H}_\texttt{k}\otimes\mathbbm{H}_\texttt{l}}. \label{CNOT}
\end{eqnarray}
Applying the encoder rules of section \ref{encoder.rules} to the encoding \eqref{CNOT}, we get the following encoder

\begin{eqnarray}
&& \mathcal{E}\left(\texttt{C-NOT}_{\texttt{k},\texttt{l}}\right)=\frac{1024}{(\lambda^3+16\lambda)^2}\left(\sigma_{k,j}^-\,\sigma_{k,j+1}^+\,\sigma_{l,j}^+
\,\sigma_{l,j+1}^- \right. \nonumber \\
&& -\sigma_{k,j}^-\,\sigma_{k,j+1}^+\,
\sigma_{l,j}^-\,\sigma_{l,j+1}^+-\sigma_{k,j}^+\,\sigma_{k,j+1}^-\,\sigma_{l,j}^-\,\sigma_{l,j+1}^- \nonumber \\
&& \left. +\sigma_{k,j}^+\,\sigma_{k,j+1}^-\,
\sigma_{l,j}^+\,\sigma_{l,j+1}^+\right) \label{CNOT-enc}
\end{eqnarray}
which is not normal. This is not a problem for the implementation of logical gates in the auxiliary space. Indeed, since single qubit unitaries and the controlled-Z form a universal set of gates, they can be multiplied in order to have another representation of the controlled-NOT gate:

\begin{equation}
\widetilde{\texttt{C-NOT}}_{\texttt{k},\texttt{l}}=\texttt{1}_{\texttt{k}}\otimes e^{-i\frac{\pi}{4}\texttt{Y}_{\texttt{l}}}\cdot\texttt{C-Z}_{\texttt{k},\texttt{l}}\cdot\texttt{1}_{\texttt{k}}\otimes e^{i\frac{\pi}{4}\texttt{Y}_{\texttt{l}}}.
\end{equation}
Due to the multiplicative properties of the map $\mathcal{E}$, the encoder of this alternative representation

\begin{equation}
\mathcal{E}\left(\widetilde{\texttt{C-NOT}}_{\texttt{k},\texttt{l}}\right)=
\mathcal{E}\left(\texttt{1}_{\texttt{k}}\right)\mathcal{E}\left(e^{-i\frac{\pi}{4}\texttt{Y}_{\texttt{l}}}\right)\mathcal{E}\left(\texttt{C-Z}_{\texttt{k},\texttt{l}}\right)\mathcal{E}\left(\texttt{1}_{\texttt{k}}\right)\mathcal{E}
\left(e^{i\frac{\pi}{4}\texttt{Y}_{\texttt{l}}}\right)
\end{equation}
is normal, being the tensor product of normal matrices. Note that the encoder $\mathcal{E}\big(\texttt{C-NOT}_{\texttt{k},\texttt{l}}\big)$ involves only four physical spins: two belonging to the $k$-th chain and two in the $l$-th chain. Nevertheless, the alternative encoder $\mathcal{E}\big(\widetilde{\texttt{C-NOT}}_{\texttt{k},\texttt{l}}\big)$ involves twelve physical spins: two in the $k$-th chain for each of the two factors $\mathcal{E}\big(\texttt{1}_{\texttt{k}}\big)$, two in the $l$-th chain for each of $\mathcal{E}\big(e^{\pm i\frac{\pi}{4}\texttt{Y}_{\texttt{l}}}\big)$, two additional spins in the $k$-th chain and two in the $l$-th chain for the encoder $\mathcal{E}\big(\texttt{C-Z}_{\texttt{k},\texttt{l}}\big)$. The increase of employed physical spins is the price to be paid for a normal encoder of the controlled-NOT gate. This does not undermine the efficiency of the algorithm because the alternative representation $\widetilde{\texttt{C-NOT}}_{\texttt{k},\texttt{l}}$ is the multiplication of a constant (in $n$) number of gates, and thus the number of physical spins in its encoder $\mathcal{E}\big(\widetilde{\texttt{C-NOT}}_{\texttt{k},\texttt{l}}\big)$ is constant in $n$.

The encoder of the operation $\texttt{A}$ \eqref{Aop} is derived from the linearity and the multiplicative properties of the map $\mathcal{E}$:

\begin{equation}
\mathcal{E}(\texttt{A}_{\texttt{k},\texttt{l}})=\mathcal{E}(|\texttt{0}_\texttt{k}\rangle\langle\texttt{0}_\texttt{k}|)\mathcal{E}(|\texttt{0}_\texttt{l}\rangle\langle\texttt{0}_\texttt{l}|)+\mathcal{E}(|\texttt{1}_\texttt{k}\rangle\langle\texttt{1}_\texttt{k}|)\mathcal{E}(|\texttt{1}_\texttt{l}\rangle\langle\texttt{1}_\texttt{l}|).
\end{equation}

{\bf Acknowledgements.} U.M. warmly thanks Andreas Winter for useful discussions. The work has been supported by grants P1-0044 and J1-5439 of Slovenian Research Agency.



%

\end{document}